\documentstyle[spie,12pt]{article}

\title{Decoherence and measurement in\\{}open quantum systems}

\author{{\Large Vladimir Privman and Dima Mozyrsky}
\skiplinehalf
Department of Physics, Clarkson University, Potsdam, New York\ 13699--5820
}

\begin{document}
\maketitle

\begin{abstract}
We review results of a recently developed model of a microscopic 
quantum system interacting with the macroscopic 
world components which are modeled by collections
of bosonic modes. The interaction is via a general operator $\Lambda$ 
of the system, coupled to the creation and annihilation 
operators of the environment modes. We assume that in the process of
a nearly instantaneous quantum measurement, the function of the environment 
involves two distinct parts: the
pointer and the bath. Interaction of the system with the bath leads to 
decoherence such that the system and the pointer both evolve into a statistical
mixture state described by the density matrix such that the system is in one
of the eigenstates of $\Lambda$ with the correct quantum mechanical probability,
whereas the expectation values of pointer operators retain amplified 
information on that eigenstate. We argue that this process represents the initial 
step of a quantum measurement. 
Calculation of the elements of the reduced
density matrix of the system and pointer is carried out exactly, and time
dependence of decoherence is identified. We discuss general implications of our model
of energy-conserving coupling to a heat bath for processes of adiabatic quantum 
decoherence. We also evaluate changes in the expectation
values of certain pointer operators and suggest that these can be interpreted as 
macroscopic indicators of the measurement outcome. 
 
\end{abstract}


\section{INTRODUCTION}

Quantum decoherence and measurement have attracted new interest owing to advances in nano\-technology. Here we present our recent work culminating in a solvable model of a measurement process.\cite{1} It has been argued that interaction with external environment, modeled by a large, macroscopic ``bath,'' is an essential ingredient of the measurement
process. Coupling to the bath is responsible for decoherence which creates a statistical mixture of eigenstates from the
initially fully or partially coherent quantum state of
the measured system. Another important macroscopic part of a measurement setup
is the ``pointer'' that stores the outcome. It has been conjectured that the bath
also plays a crucial role
in the selection of those quantum states of the pointer
that manifest themselves in classical observations.\cite{2,3,4,5,6,7}
In this work we present a model in which a multimode pointer retains 
information on the measurement outcome because of its 
coupling to the measured system, without the need to couple it also
directly to the bath. However, the measured system is assumed to be coupled to the bath.
The latter then also affects the dynamics of the pointer because both of them are coupled 
to the measured system.

Exact results for models of quantum systems interacting with their environment are quite limited. In the framework of an exactly solvable model of a quantum oscillator coupled
to a heat bath of oscillators, it has been shown\cite{4} that the
reduced density matrix of the system looses its off-diagonal elements in the
eigenbasis of the interaction Hamiltonian. Such loss of correlations between the states of a quantum system is a manifestation of decoherence. Recent work on
decoherence\cite{8,9,10,11} has explored its 
effects for rather general cases, for bosonic (oscillator) and 
spin baths. Applications for various physical systems have been
reported.\cite{12,13,14,15,16,17,18}
Fermionic heat bath has also been considered.\cite{19}

We note that a true heat bath should cause a microscopic system interacting with it to
thermalize. In fact, we are not aware of any adequate general, truly microscopic model of such a process. Instead, the ``temperature'' is usually introduced phenomenologically via the initial state of the noninteracting bath modes. The function of a measuring device
is different from and more complex than that of a heat bath. In particular, it must store and amplify the measurement outcome information. 
One of the key issues
in the description of a measurement process is the
interpretation of the transfer of information stored after the 
system-pointer and system-bath interaction to the 
macroscopic level.\cite{2} Here we offer a model of the process which corresponds to the first stage of measurement, in which the pointer acquires amplified information by
entanglement with the state of the system. Thus we do not claim to resolve
the foundation-of-quantum-mechanics issue of how that information is 
passed on to the classical world, involving the collapse of the wave functions of the system and of each pointer mode. Indeed, it is impossible to fully describe 
the wave function collapse within the unitary quantum-mechanical description of the three systems involved: the measured system, the pointer, and the bath, the latter being internal in the sense
that, in our model, it only interacts with the measured system. One would have to 
consider an external bath (the rest of the universe) with which all the ``internal'' systems interact. As far as we know, this problem is not presently solved,\cite{2,3} and we first sidestep it by assuming separation of time scales (see below). Nevertheless, we later argue that the results of our model provide a useful insight into this aspect of the quantum measurement process.

Let us identify the three quantum systems involved. 
The first one, that is being measured, $S$, is a microscopic system with the
Hamiltonian which will be also denoted by $S$.
Second, the measuring device must
have the ``bath'' or ``body'' part, $B$, containing many 
modes. The $k$th mode will have the Hamiltonian $B_k$. We assume that the
bath modes are not coupled to each other.
The bath part of the measuring device is not observed, i.e., it can be
traced over.
The last system is the pointer, $P$, consisting of many modes
(that are not traced over). The pointer
amplifies the information obtained in the measurement
process and can later pass it on for further amplification
or directly to macroscopic (classical) systems.
The $m$th pointer mode has the Hamiltonians $P_m$. It is assumed that
expectation values of some quantities in the pointer undergo 
a large change during the measurement.

It will become evident later that the device modes involved in the 
measurement process can be rather simple so that one can focus on the
evolution of the system $S$ and its effect on the pointer $P$.
However, it is the pointer interaction with the external
bath (some external modes, ``the rest of the universe'')
that is presumed to select those quantum states of $P$
that manifest themselves classically. For now, we prefer to avoid the
discussion of this matter,\cite{2,3,4,5,6} by assuming that
the added evolution of the pointer due to such external interactions
occurs on time scales larger than the measurement time, $t$. Similarly,
when we state that the internal bath modes can be ``traced over,'' we 
really mean that their interactions with the rest of the universe are such 
that these modes play no role in the later, wave-function-collapse stage of the 
measurement process.

Moreover, the measuring device probes
the state of the system $S$ at the initial time, $t=0$,
rather than its time evolution under $S$ alone. Ideally the process of measurement is 
instantaneous. In practice, it has to be faster than the time scales
associated with the dynamics under $S$ and the evolution due to the 
interactions of all the three systems involved, $S,B,P$, with the rest of the universe. This
can be obtained as the limit of a system in which very strong
interactions between $S$ and $B$, and also between $S$ 
and $P$, are switched on at $t=0$ and switched off at $t>0$,
with small time interval $t$. Of course, at later times the pointer can 
interact with other, external systems to pass on
the result of the measurement. Thus, in our approach, we will assume that the 
Hamiltonian of the system itself, $S$, as well as all the external interactions,
can be neglected for the duration of the measurement, $t$.

In the next section we introduce our model along the lines outlined
above. In Section\ 3, we carry out exact coherent-state calculations that explicitly 
show how the system and the pointer evolve into a statistical mixture of direct 
product states, representing the required framework for quantum measurement.
In Section\ 4, we describe the emergence of decoherence in the continuum limit of an
infinite number of modes. We also present a general discussion of adiabatic quantum decoherence.\cite{8} Finally, in the last section we 
calculate expectation values of some pointer operators that retain information on the outcome of the
measurement process, and discuss implications of our results. 

\section{THE MODEL}

The total Hamiltonian
of the system and the measuring device will be written
based on the assumptions presented in the preceding section.
Specifically, the internal Hamiltonian of the measured system
is ignored (set to a constant which can be zero without loss of generality).
This is because the dynamics of measurement is assumed to occur on the
time scale $t$ much shorter than any internal dynamical evolution of the
measured system, in order to probe only the instantaneous 
system wavefunction
in the measurement process. We take

$$ H=\sum_k B_k + \sum_m P_m + b \Lambda 
\sum_k {\cal B}_k +  p \Lambda \sum_m {\cal P}_m  \; . \eqno(1) $$

\noindent{}Here $\Lambda$ is some Hermitian operator of the system
that couples to certain operators of the modes, ${\cal B}_k$
and ${\cal P}_m$. The subscript $k$ labels the noninteracting (with each other) modes of the
bath, with their Hamiltonians $B_k$, whereas $m$ labels similar modes of the pointer, with Hamiltonians $P_m$.
The parameters $b$ and $p$ are
introduced to measure the coupling strength for the
bath and pointer modes to the measured system, respectively. 
They are assumed very large;
the ideal measurement process could correspond to $b,p\to \infty$.

We note that the modes of $P$ and $B$ can be similar. The only difference between the bath and pointer modes is in how they interact with the ``rest of the universe'' in a later stage of the measurement process: the bath is not observed (traced over), whereas the pointer modes have their wave functions collapsed. Through their entanglement with the measured system in the first stage of the measurement process, treated here, the pointer modes also cause the collapse of the system wavefunction. Thus, we actually took the same coupling operator $\Lambda$ for the bath and  pointer. In fact, all the exact calculations reported
in this work can be also carried out for different coupling
operators $\Lambda_B$ and $\Lambda_P$, for the bath and pointer
modes, provided they commute, $[\Lambda_B,\Lambda_P]=0$, so that
they share a common set of eigenfunctions. The final wavefunction
of the measured system, after the full measurement, is in this set. 
Analytical calculation can be even extended to
the case when the system Hamiltonian $S$ is retained in (1),
provided all three operators, $S,\Lambda_B,\Lambda_P$, 
commute pairwise. The essential physical ingredients of the model
are captured by the simpler choice (1).

We will later specify all the operators in (1) as
the modes of the bosonic heat bath of Caldeira-Leggett
type.\cite{17,19,20,21,22,23,24,25,26} For now, however, let us keep our discussion
general. We will assume that the system operator $\Lambda$ 
has a nondegenerate, discrete spectrum of 
eigenstates:

$$ \Lambda | \lambda \rangle = \lambda | \lambda \rangle \; . \eqno(2) $$

\noindent{}Some additional
assumptions on the spectrum of $\Lambda$ and $S$ will be 
encountered later. We also note that the requirement that the coupling parameters
$b$ and $p$ are large may in practice be satisfied by that, at the time
of the measurement, the
system Hamiltonian $S$ corresponds to slow or trivial dynamics.

Initially, at $t=0$,
the quantum systems $S,B,P$ and their modes
are not correlated with each other.
We assume that $\rho$ is the initial density matrix of 
the measured system. The initial state of each bath and pointer
mode will be assumed thermalized, with $\beta=1/(kT)$ and
the density matrices

$$ \theta_k = {e^{-\beta B_k}\over{\rm Tr}_k \left(e^{-\beta B_k}\right)} \; ,
\qquad\qquad
\sigma_m = {e^{-\beta P_m}\over{\rm Tr}_m \left(e^{-\beta P_m}\right)} \; , \eqno(3) $$

\noindent{}respectively. We cannot offer any fundamental physical reason for having the initial
bath and pointer mode states thermalized, especially for the pointer. 
This choice is really made to allow exact solvability, though we could claim that the bath and pointer
might be thermalized if they are in contact with the ``rest of the universe'' for a long time before the measurement.

The 
density matrix of the full system at time $t$ 
is then

$$ R=e^{-iHt/\hbar}\left[ \rho\left(\prod_k \theta_k\right)\left( 
\prod_m \sigma_m\right) \right] e^{iHt/\hbar} \; . \eqno(4) $$

\noindent{}The bath is not probed and it can be traced over.
The resulting reduced density matrix $r$ 
of the combined system $S+P$ will be
represented as by its matrix elements in the eigenbasis
of $\Lambda$. These
quantities are each an operator in the space 
of $P$:

$$ r_{\lambda\lambda^\prime}=\langle 
\lambda|{\rm Tr}_B(R)|\lambda^\prime\rangle \; . \eqno(5)$$

We now assume that operators in different spaces and of different
modes commute. Note that until now our discussion was quite general.
The commutability requirement is trivial for the bosonic and spin
bath modes. However, it must be checked carefully if baths with fermionic modes are used. Then one can show that

$$ r_{\lambda\lambda^\prime}=\rho_{\lambda\lambda^\prime} \left[ \prod_m
e^{-i t \,\left( P_m + p \lambda {\cal P}_m \right)/\hbar} \sigma_m 
e^{i t \,\left( P_m + p \lambda^\prime {\cal P}_m \right)/\hbar} \right] 
\left[ \prod_k {\rm Tr}_k \left\{ e^{-i t \,\left( B_k + b \lambda {\cal B}_k \right)/\hbar} \theta_k 
e^{i t \,\left( B_k + b \lambda^\prime {\cal B}_k \right)/\hbar} \right\} \right] \; , \eqno(6) $$

\noindent{}where 
$\rho_{\lambda\lambda^\prime}=\langle 
\lambda| \rho |\lambda^\prime\rangle$.
This result involves products of $P$-space operators
and traces over $B$-space operators which are all single-mode. Therefore,
analytical calculations are possible for some choices of the
Hamiltonian (1). The observable $\Lambda$ can be kept general.

The role of the product of traces over the modes of the bath 
in (6) is to induce decoherence
which is recognized as essential for the 
measurement process.\cite{2,3,4,5,6,7,8} At time $t$, the absolute
value of this product should approach $\delta_{\lambda\lambda^\prime}$
in the limit of large $b$. Let us now assume that the bath is 
bosonic. The Hamiltonian
of each mode is then $ \hbar\omega_k a^\dagger_k a_k$, where
for simplicity we shifted the zero of the oscillator energy to the
ground state. The coupling operator ${\cal B}_k$ is usually selected as
$g^*_ka_k + g_k a_k^\dagger$. For simplicity, though,
we will assume that the coefficients $g_k$ are 
real:

$$B_k = \hbar \omega_k a^\dagger_k a_k \; ,
\qquad\quad\quad 
{\cal B}_k=g_k\left(a_k + a_k^\dagger\right) \; . \eqno(7)$$

\noindent{}For example, for radiation field in a unit volume,
coupled to an atom,\cite{27}
the coupling is via a linear combination of the operators
$(a_k + a_k^\dagger )/\sqrt{\omega_k}$ and 
$i(a_k - a_k^\dagger )/\sqrt{\omega_k}$. For a spatial oscillator,
these are proportional to position and momentum, respectively. 
Our calculations can be extended to have an
imaginary part of $g_k$ which adds interaction with momentum.

\section{COHERENT STATE CALCULATION}
 
In order to calculate traces in (6), we utilize the coherent-state formalism.\cite{27}
The coherent states $|z\rangle$ are
the eigenstates of the annihilation
operator $a$ with complex eigenvalues $z$. 
Note that from now on we omit the oscillator
index $k$ whenever this leads to no confusion. 
These states are not orthogonal:

$$ \langle z_1|z_2\rangle =\exp{\left( z_1^*z_2-{1\over 2}|z_1|^2-
{1\over 2}|z_2|^2\right)} \; . \eqno(8) $$   

\noindent{}They form an over-complete set, and one can
show that the identity operator
in a single-oscillator space can be obtained as the integral

$$\int d^2z \, |z\rangle\langle z|=1 \; . \eqno(9) $$

\noindent{}Here the integration by definition corresponds to

$$ d^2z \equiv {1\over \pi}\, d\left({\rm Re}z\right) 
d\left({\rm Im}z\right)  \; . \eqno(10) $$

\noindent{}Furthermore, for an arbitrary operator $A$, we have,
in a single-oscillator space, 

$$ {\rm Tr}\, A = \int d^2z \, \langle z|A|z\rangle  \; . \eqno(11) $$

\noindent Finally, we note the
following identity,\cite{27} which will be used later,

$$e^{\Omega a^{\dag}a}={\cal
N}\left[e^{a^{\dag}(e^{\Omega}-1)a}\right]  \; . \eqno(12) $$

\noindent{}In this relation $\Omega$ is an arbitrary c-number, while ${\cal N}$ 
denotes normal ordering.

It is convenient to introduce operators $\gamma_{\lambda}$ according to

$$\hbar\gamma_{{\lambda},k}=B_k + \lambda b{\cal B}_k  \; . \eqno (13)$$

\noindent{}Denoting the $k$th term in the second product in (6) by $U_{\lambda\lambda^{\prime},k}$
and utilizing relations (8)-(12), we have

$$ U_{\lambda\lambda^{\prime}}=Z^{-1}\int d^2z_0\, d^2z_1\ d^2z_2 \ \langle
z_0|e^{-it\,\gamma_{\lambda}}|z_1\rangle
\langle z_1|e^{-\hbar\beta\omega a^{\dag} a}|z_2\rangle \langle z_2|
e^{it\,\gamma_{\lambda^{\prime}}}|z_0\rangle  \; . \eqno (14)$$

\noindent{}Here we again omitted the mode index $k$ for simplicity, and $Z$ stands for $Z_k \equiv {\rm Tr}_k \left(e^{-\beta B_k}\right)$.
The normal-ordering formula (12) then yields for the middle term,

$$\langle z_1|e^{-\hbar\beta\omega a^{\dag} a}|z_2\rangle = 
\langle z_1|z_2\rangle e^{z_1^*(e^{-\hbar\beta\omega}-1)z_2}=\exp{\left[z_1^*z_2-{1\over
2}|z_1|^2-{1\over 2}|z_2|^2 +z_1^*(e^{-\hbar\beta 
\omega} -1)z_2\right]}  \; . \eqno (15)$$

In order to evaluate the first and last matrix-element factors in (14), we define
the shifted operators

$$ \eta = a + \lambda b(\hbar\omega)^{-1}g  \; , \eqno (16)$$

\noindent{}in terms of which we have

$$\gamma_{\lambda}=\omega\eta^{\dag}\eta- \lambda^2 b^2(\hbar^2\omega)^{-1}g^2  \; . \eqno (17)$$

\noindent{}Since $\eta$ and $\eta^{\dag}$ still satisfy the bosonic
commutation relation $[\eta,\eta^{\dag}]=1$, the normal-ordering
formula applies. Thus, for the first matrix element in (14), for instance,
we get

$$\langle z_0|e^{-it\gamma_{\lambda}}|z_1\rangle=e^{it\lambda^2 b^2
{g^2/( \hbar^2\omega)}}\langle z_0|z_1\rangle e^{
(e^{-i\omega t}-1) [z_0^*+\lambda b
{g/(\hbar\omega)}] [z_1+\lambda b{g/(\hbar\omega)}]}  \; . \eqno (18)$$

Collecting all the results, one concludes that
the calculation of $U_{\lambda\lambda^{\prime}}$ involves six Gaussian integrations
over the real and imaginary parts of the variables $z_0, z_1,
z_2$. This is a rather lengthy calculation but it can be carried out
in closed form. The result, with indices $k$ restored, is

$$  \left| \, \prod_k U_{\lambda\lambda^{\prime},k} \, \right| = \exp\left[-2b^2
\left(\lambda-\lambda^\prime\right)^2 \Gamma (t) \right] \; , \eqno(19)$$

\noindent{}with

$$\Gamma(t)= \sum_k (\hbar\omega_k)^{-2} g_k^2\sin^2 {\omega_k t\over 2}
\coth{{\hbar\beta\omega_k \over 2}}  \; . \eqno(20) $$

\noindent{}We only gave the expression for the absolute value of the product of traces in (6). Its complex phase is also known\cite{8} but plays no role in our considerations. In the next section we analyze the results (19)-(20) in the continuum limit of a large number of modes.

\section{ADIABATIC DECOHERENCE}

Results of the preceding section suggest that the off-diagonal elements of the effective density matrix, obtained after tracing over a bath, are generally decreased by the interaction with the bath modes. We will shortly demonstrate that in the limit of many bath modes, these off-diagonal elements irreversibly decay, thus leaving the system and the pointer in a statistical mixture of direct-product states. This process is usually termed decoherence. Apart from being of fundamental importance in the theory of quantum measurement, decoherence has attracted much interest recently due to rapid development of new fields such as quantum computing and quantum information.\cite{8,11,28,29} Decoherence due to external interactions is a
major obstacle to implementation of coherently evolving quantum devices such
as quantum computers. In this section we review some aspects of the physics of decoherence.\cite{8}

Decoherence is a result of the coupling of the quantum system to the environment which, generally, is the rest of the
universe. In various experimentally relevant situations interaction
of a quantum system with environment is dominated/mediated by its 
microscopic surroundings that can be represented by a set of modes. For example, the dominant source of such
interaction for an atom in an electromagnetic cavity field is the
field itself coupled to the dipole moment of the
atom.\cite{27} In the case of Josephson junction in a magnetic
flux\cite{21,22} or defect propagation in solids, the interaction can be
dominated by acoustic phonons or delocalized electrons.
Magnetic macromolecules interact with the surrounding spin environment
such as nuclear spins,\cite{12} etc.

It is customary\cite{23,25} to model a heat bath as a system of noninteracting boson modes. It has been established, for harmonic
quantum systems,
that the influence of the heat bath described by the oscillators is
effectively identical to an external uncorrelated random force acting
on a quantum system under consideration. In order for the system to
satisfy equation of motion with a linear dissipation term, in the
classical limit, the coupling was chosen to be linear in coordinates
while the coupling constants entered lumped in a spectral function, such as (20),
and were assumed to be of a power-law form in the oscillator
frequency, with the appropriate Debye cutoff. 

The temperature was introduced via the initial thermal state of the bath, as in
(3).
This model of a heat bath was applied to study effects of
dissipation on the probability of quantum tunneling from a metastable
state.\cite{20} It was found that coupling a quantum 
system to the heat bath actually decreased the quantum tunneling rate.
The problem of a particle in a double well potential was also
considered.\cite{21,22}  
In this case the interaction with the bath lead to 
quantum decoherence and complete localization at zero temperature.
This study has lead to the spin-boson Hamiltonian\cite{21,22} which
found numerous other applications.

For a general system, there is no systematic way to separate decoherence and
thermalization effects.
We note that thermalization is naturally associated with exchange of
energy between the quantum system and heat bath.
Model system results and general expectations mentioned
earlier suggest that in many cases decoherence
involves its own time scales which are shorter than those of approach
to thermal equilibrium.\cite{3,4,5,8} Our measurement model 
represents, in fact, a limiting case such that there is no energy exchange 
between the system and the bath. This situation essentially corresponds to
the early stages of the system-bath interaction, at low temperatures, when thermalization 
effects are not yet significant. 

Thus if we ignore the presence of the pointer (set $p=0$), and restore the system Hamiltonian $S$ in (1), then
our model of the system-bath interaction, with $ H=S+\sum_k B_k +  \Lambda 
\sum_k {\cal B}_k $, now with $b=1$ but still with {\it commuting\/} $S$ and $\Lambda$,
corresponds to adiabatic quantum decoherence.\cite{8}
Indeed, the commutation property means that $[H,S]=0$, so that the system energy is conserved and therefore relaxation is generally suppressed. Only decoherence processes are possible and in fact exactly calculable.\cite{8,9,10,11}
The results (19)-(20) remain unchanged and indicate that decoherence is controlled by the interaction with
the bath, $\Lambda$, rather than by the system Hamiltonian. The eigenvalues of the ``pointer observable'' $\Lambda$ determine the rate of decoherence, while the type of the bath and coupling determines the function $ \Gamma (t) $.

We also note that $ \Gamma (t) $ is a sum of positive terms. For true decoherence, i.e., in order for this sum to diverge for {\it large\/} times, one needs a continuum of frequencies and interactions with the bath modes that are
strong enough at low frequencies.\cite{8} We will give specific results shortly. For quantum measurement, thought, the time $t$ need not be large. The conditions for decoherence are then somewhat different; see below.

In the continuum limit of many modes, the density of the bosonic 
bath states in unit volume,
${\cal D}(\omega)$, and the Debye cutoff, with frequency $\omega_D$,
are introduced\cite{22} so that (20) can be rewritten as

$$ \Gamma(t)= \int\limits_0^\infty \, d\omega \,   
{{\cal D}(\omega) g^2(\omega)\over(\hbar\omega)^2}\, 
e^{-\omega/\omega_D}\, \sin^2 {\omega t\over 2}
\coth{{\hbar\beta\omega \over 2}} \; . \eqno(21) $$

\noindent{}Let us consider the usual choice,\cite{8} motivated by atomic-physics\cite{27} and
solid-state applications,\cite{22} corresponding to

$$ {\cal D}(\omega) g^2(\omega) = \Omega \, \omega^n \; , \eqno(22)$$
where $\Omega$ is a constant. 

Exponent values $n=1$ and $n=3$ have been analyzed in detail in the literature.\cite{8,9,10,11,22} 
For $n=1$, three regimes were identified, defined by the time scale for thermal decoherence, $\hbar\beta$, which is large for low temperatures,  and the time scale
for quantum-fluctuation effects, $\omega_D^{-1}$. The first,\cite{11} 
``quiet'' regime, $t \ll \omega_D^{-1}$,
corresponds to no significant decoherence and $\Gamma \propto 
(\omega_D t)^2$. The next,
``quantum'' regime, $ \omega_D^{-1} \ll t \ll \beta$,
corresponds to decoherence driven
by quantum fluctuations and $\Gamma \propto \ln (\omega_D t)$.
Finally, for $t \gg \beta$, in the ``thermal'' regime,
thermal  fluctuations are dominant and $\Gamma \propto
t/\beta $.
For $n=3$, decoherence is incomplete.\cite{11} Indeed, while $n$
must be positive for the integral in (21) to converge, only for
$n<2$ we have divergent $\Gamma (t)$ growing according to a power
law for large times (in fact, $\Gamma \propto t^{2-n}$) 
in the ``thermal'' regime. Thus, strong enough coupling $|g(\omega)|$
to the low-frequency modes of the heat bath is crucial for full decoherence.

Let us concentrate on the $n=1$ case, motivated by atomic physics\cite{27} and solid-state applications;\cite{22} it is termed Ohmic dissipation. We will now consider conditions for decoherence required for quantum measurement. These are somewhat different from those just discussed for the large-time adiabatic system-bath interaction effects. Here we assume
that the relevant energy gaps of $S$ are bounded so that there
exists a well defined time scale $\hbar/\Delta S $ of the evolution
of the system under $S$. There is also
the time scale $1/\omega_D$ set by the frequency cutoff assumed
for the interactions. The thermal time scale is $\hbar \beta$. 
The only real limitation on the duration of measurement
is that $t$ must be less then $\hbar/\Delta S$. In applications,
typically\cite{22} one can assume that $1/\omega_D \ll \hbar/\Delta S$.
Furthermore, it is customary to assume that the temperature is 
low,\cite{8,22}
 
$$ t {\rm\ and\ } 1/\omega_D \ll \hbar/\Delta S \ll \hbar\beta  \; . \eqno(23) $$

\noindent{}In 
the limit of large $\hbar\beta$, for Ohmic dissipation, (19) reduces 
to

$$ \left| \prod_k U_{\lambda\lambda^{\prime},k} \right| \simeq
 \exp\left\{ -{\Omega
\over 2 \hbar^2} b^2
\left(\lambda-\lambda^\prime\right)^2 \ln \left[1+(\omega_D t)^2\right]
\right\} \; . \eqno(24) $$

\noindent{}In 
order to achieve effective decoherence, the product 
$ b^2(\Delta \lambda)^2 \ln [1+(\omega_D t)^2] $
must be large. The present 
approach only applies to operators $\Lambda$
with nonzero scale of the smallest spectral gaps, $\Delta \lambda$.

We thus note that, unlike the large-time system-bath decoherence, the decoherence property needed for
the measurement process will be obtained for nearly any well-behaved
choice of ${\cal D}(\omega)g^2(\omega)$ because we can rely on the value
of $b$ being large rather than on the properties of
the function $\Gamma (t)$, which can no longer be considered evaluated at large $t$. If $b$ can be large enough,
very short measurement times $t$ are possible. However, it may be advisable
to use measurement times $ 1/\omega_D \ll t \ll \hbar/\Delta S$ to 
get the extra amplification factor $\sim \ln (\omega_D t)$ and allow
for fuller decoherence and less sensitivity to the value of $t$ in the
pointer part of the dynamics, to be addressed in the next section.
We notice, furthermore, that the assumption
of a large number of modes is important for the monotonic decay
of the absolute value (19) in the ``system-bath'' decoherence applications,\cite{8,9,10,11} where irreversibility
is obtained only in the limit of the infinite number of modes. In the measurement case, it
can be shown\cite{1} that the role of
such a continuum limit is mainly to allow to 
extend the possible measurement times
from $t \ll 1/\omega_D$ to $1/\omega_D \ll t \ll \hbar/\Delta S$.

\section{POINTER PROPERTIES, AND DISCUSSION}
 
Consider the reduced density matrix
$r$ of $S+P$, with matrix elements given by (6). It becomes diagonal in $|\lambda\rangle$,
at time $t$, because all the nondiagonal elements are  
small, assuming that effective decoherence has been achieved, as discussed in the preceding section:

$$ r=\sum_\lambda |\lambda \rangle \langle \lambda | \, \rho_{\lambda\lambda}
\prod_m e^{-i t \left( P_m + p \lambda {\cal P}_m \right)/\hbar} \sigma_m 
e^{i t \left( P_m + p \lambda {\cal P}_m \right)/\hbar} \; . \eqno(25)$$

\noindent{}Thus, the dynamics yields the density matrix that can be interpreted as describing a statistically distributed
system (a mixture), without quantum correlations. This, however, is only meaningful within the 
ensemble interpretation of quantum mechanics. 
For a single system plus device, coupling to the rest of the universe is presumably needed to describe how that system is left in one of the eigenstates $|\lambda \rangle$,
with probability $\rho_{\lambda\lambda}$. Presently, such a process of wavefunction collapse is not fully understood,\cite{2} but see comments below on the implications of our model for this problem. After the
measurement interaction is switched off at $t$, the pointer
of that system will carry information on the value of $\lambda$. 
This information is ``amplified,''
owing to the large parameter $p$ in the interaction. 

We note that one of the roles of the pointer having many modes, which can be identical and noninteracting, is to allow it (the pointer only) to still be treated in the ensemble, density matrix description, even if we focus on the late stage of the measurement when the wave functions of a single measured system and of each pointer mode are already collapsed. This pointer density matrix can be read off (25); it is the $\lambda$-dependent product over the pointer modes labeled by $m$. 

Another useful insight is provided by the fact that, as will be shown shortly, the changes in the expectation values of some observables of the pointer retain amplified information on the system eigenstate. The coupling to the rest of the universe that leads to the completion of the measurement process, should involve such an observable of the pointer. Eventually, the information in the pointer, perhaps after several steps of amplification,
should be available for probe by interactions with classical devices. 

At time $t=0$, expectation values of various operators 
of the pointer will have their initial values. These values will
be changed at time $t$ of the measurement owing to the
interaction with the measured system. It is expected that
the large coupling parameter $p$ will yield large 
changes in expectation values of the pointer quantities. 
This does not apply equally to
all operators in the $P$-space. Let us begin with the simplest choice: 
the Hamiltonian $\sum\limits_m P_m$
of the pointer. 

We will assume that the pointer is described by
the bosonic heat bath and, for simplicity, use the same notation
for the pointer modes as that used for the bath modes. The assumption
that the pointer modes are initially thermalized, see (3),
was not used thus far.
While it allows exact analytical calculations, it is not essential: the
effective density matrix describing the pointer modes at time $t$, for the system state $\lambda$, will retain amplified information on the value of $\lambda$ for general initial states of the pointer. 

This effective density matrix is the product over the $P$-modes in (25). For the ``thermal'' $\sigma_m$ from (3), the
expectation value of the pointer energy, $\langle E_P \rangle_\lambda$, can be calculated 
from

$$ {{\rm Tr}_P \left\{ \left( \sum_m \hbar\omega_m a^\dagger_m a_m 
\right) \prod_n \left[ e^{-i t [  \omega_n a^\dagger_n a_n + p \lambda 
g_n(a_n + a_n^\dagger) /\hbar]} \left(e^{-\hbar\beta \sum_k 
\omega_k a^\dagger_k a_k}\right) e^{i t [ \omega_n a^\dagger_n a_n + p \lambda 
g_n(a_n + a_n^\dagger) /\hbar]}\right] \right\} \over {\rm Tr}_P \left(e^{-\hbar\beta \sum_s 
\omega_s a^\dagger_s a_s }\right)} \; . \eqno(26)$$

\noindent{}This expression can be reduced to calculations for individual modes.
Operator identities can be then utilized to obtain, after some lengthy algebra, the 
results

$$ \langle E_P \rangle_\lambda (t) = \langle E_P 
\rangle (0) + \langle \Delta E_P \rangle_\lambda (t) \; , \eqno(27)$$

$$ \langle E_P \rangle (0) = \hbar \sum_m \omega_m e^{-\hbar\beta \omega_m} \left(1-
e^{-\hbar\beta \omega_m}\right)^{-2} \; , \eqno(28) $$

$$\langle \Delta E_P \rangle_\lambda (t)=  {4 p^2 \lambda^2 \over \hbar} 
\sum_m {g_m^2 \over 
\omega_m} \sin^2 \left({\omega_m t \over 2}\right) \; . \eqno(29) $$

\noindent{}For a model with Ohmic dissipation, the integral in the
continuum limit can be calculated to 
yield

$$ \langle \Delta E_P \rangle_\lambda (t)=  {2\, 
\Omega \,\omega_D \lambda^2 p^2 \over \hbar}
{(\omega_Dt)^2 \over 1 + (\omega_Dt)^2} \; . \eqno(30)$$

\noindent{}The energy will be an indicator
of the amplified value of the square of $\lambda$, provided $p$
is large. Furthermore, we see here the advantage of
larger measurement times, $t \gg 1/\omega_D$. The change in the
energy then reaches saturation. After time $t$, when the
interaction is switched off, the energy of the pointer
will be conserved, internally, i.e., until the pointer is affected by the
``rest of the universe.''

Let us consider the expectation value of the following 
Hermitian operator of the 
pointer:

$$ X=\sum_m {\cal P}_m = \sum_m g_m (a_m +a_m^\dagger) \; . \eqno(31) $$

\noindent{}For
an atom in a field, $X$ is related to the electromagnetic
field operators.\cite{24} One can show that
$\langle X_P \rangle (0) =0$ 
and

$$ \langle \Delta X_P \rangle_\lambda (t)=  \langle 
X_P \rangle_\lambda (t) = -{4 p \lambda \over \hbar} \sum_m {g_m^2 \over 
\omega_m} \sin^2 \left({\omega_m t \over 2}\right) =  -{2\, 
\Omega \,\omega_D \lambda p \over \hbar}
{(\omega_Dt)^2 \over 1 + (\omega_Dt)^2} \; . \eqno (32)$$

\noindent{}The 
change in the expectation value is linear in $\lambda$.
However, this operator is not
conserved internally. One can show that after time $t$ its expectation value 
decays to zero for times $t + {\cal O} (1/\omega_D)$.

We note that by referring to ``unit volume'' we have avoided
the discussion of the ``extensivity'' of various quantities. 
For example, the initial energy $\langle E_P 
\rangle (0)$ is obviously
proportional to the system volume, $V$. However, 
the change 
$\langle \Delta E_P \rangle_\lambda (t)$ 
will not be extensive; typically,
$g^2(\omega)\propto 1/V$, ${\cal D}(\omega) \propto V$. Thus, while the amplification
in our measurement process can involve a numerically large 
factor, the changes in the quantities
of the pointer will be multiples of microscopic values. Multi-stage
amplification, or huge coupling parameter $p$, would be needed for
the information in the pointer to become truly ``extensive'' macroscopically.

In summary, we described the first stage of a measurement process. 
It involves decoherence
due to a bath and transfer of information 
to a large system (pointer) via
strong interaction over a short period of time.
The pointer itself need not be coupled to the internal bath.
While we do not offer a solution to the foundation-of-quantum-mechanics
wave-function collapse problem,\cite{2} our results do provide two interesting observations. 

Firstly, the pointer operator ``probed'' by the rest of the universe during the wave-function collapse stage, determined by how the pointer modes are coupled to the external bath,\cite{3,4,5,6,7} must have the appropriate amplification capacity in the first, decoherence stage of the measurement process, as illustrated by our calculations.

Secondly, 
for a single system (rather then an ensemble), the multiplicity of the (noninteracting) pointer modes might allow the pointer to be treated within the density matrix formalism even after the system and each pointer-mode wave functions were collapsed. Since it is the information in the pointer that is passed on,
this observation might seem to resolve part of the measurement puzzle.
Specifically, it might suggest why only those density matrices entering 
(25) are selected for the pointer: they carry classical (large, different from other values) information in expectation values, rather than quantum-mechanical superposition.  

However, presumably\cite{2} only a full description of the interaction of the external world with the system $S+P$ can explain the wavefunction collapse.
It is likely that in practice there will be two types of pointer involved, in a multistage
measurement process. Some pointers will consist of many 
{\it noninteracting\/} modes.
These pointers carry the information, stored in a density matrix rather
than a wave function of a single system. The latter transference hopefully makes the
wavefunction collapse and transfer of the stored information to the macroscopic level less
``mysterious.'' The second type of pointer will involve 
strongly
interacting modes and play the role of an amplifier by utilizing the 
many-body
collective behavior of the coupled modes, phase-transition style. Its role will be
to alleviate the requirement for large mode-to-system coupling
parameters encountered in our model.
 
We acknowledge helpful discussions with
Professor L.~S.~Schulman. This research has been supported
by the US Army Research Office under grant 
DAAD$\,$19-99-1-0342.

{}

\frenchspacing
\begin{thebibliography}{99}

\bibitem{1} A short summary of our model of quantum measurement was presented in:\ {\it Measurement of a Quantum System Coupled to Independent Heat-Bath and Pointer Modes},
D. Mozyrsky and V. Privman, to appear in Mod. Phys. Lett. B (2000); available at
http://xxx.lanl.gov/abs/cond-mat/0003019.

\bibitem{2} J. Bell, Phys. World {\bf 3}, August 1990, No. 8, p. 33.

\bibitem{3} W. H. Zurek, Physics Today, October 1991, p. 36.

\bibitem{4} W. G. Unruh, W. H. Zurek, Phys. Rev. D {\bf 40},
1071 (1989).

\bibitem{5} W. H. Zurek, S. Habib and J. P. Paz,
Phys. Rev. Lett. {\bf 70}, 1187 (1993).

\bibitem{6} M. Gell-Mann and J. B. Hartle, in 
{\sl Proc. of the 25th International Conference on High Energy 
Physics\/} (South East Asia Theor. Phys. 
Assoc., Phys. 
Soc. of Japan, Teaneck, NJ, 1991) Vol. 2, p. 1303. 

\bibitem{7} M. Gell-Mann and J. B. Hartle, in  {\sl Quantum 
Classical Correspondence: The 4th Drexel Symp. on Quantum Nonintegrability}, ed. by D. H. Feng and B. L. Hu
(Internat. Press, Cambridge, 1997) p. 3.

\bibitem{8} D. Mozyrsky and V. Privman, J. Stat. Phys. {\bf 91},
787 (1998).

\bibitem{9} N. G. van Kampen, J. Stat. Phys. {\bf 78},
299 (1995).

\bibitem{10} J. Shao, M.-L. Ge and H. Cheng, Phys. Rev. E {\bf 53},
1243 (1996).

\bibitem{11} G. M. Palma, K. A. Suominen and A. K. Ekert,
Proc. Royal Soc. London A {\bf 452}, 567 (1996).

\bibitem{12} I. S. Tupitsyn, N. V. Prokof'ev, P. C. E. Stamp,
Int. J. Modern Phys. B {\bf 11}, 2901 (1997).

\bibitem{13} C. W. Gardiner, {\sl Handbook of Stochastic Methods
for Physics, Chemistry and the Natural Sciences\/}
(Springer-Verlag, Berlin, 1990).

\bibitem{14} A. J. Leggett, in {\sl Percolation, Localization and
Superconductivity, NATO ASI Series B: Physics}, 
ed. by A. M. Goldman and S. A. Wolf (Plenum, NY, 1984) 
Vol. 109, p. 1.

\bibitem{15} J. P. Sethna, Phys. Rev. B {\bf 24}, 698 (1981).

\bibitem{16} A. O. Caldeira and A. J. Leggett,
Ann. Phys. {\bf 149}, 374 (1983).

\bibitem{17} A. Garg, Phys. Rev. Lett. {\bf 77}, 764 (1996).

\bibitem{18} L. Mandel and E. Wolf, {\sl Optical Coherence 
and Quantum Optics\/} (Cambridge Univ. Press, 1995).

\bibitem{19} L.-D. Chang and S. Chakravarty, Phys. Rev. B
{\bf 31}, 154 (1985).

\bibitem{20} A. O. Caldeira and A. J. Leggett,
Phys. Rev. Lett. {\bf 46}, 211 (1981).

\bibitem{21} S. Chakravarty and A. J. Leggett, Phys. Rev. Lett.
{\bf 52}, 5 (1984).

\bibitem{22} A. J. Leggett, S. Chakravarty, A. T. Dorsey,
M. P. A. Fisher and W. Zwerger, Rev. Mod. Phys. {\bf 59}, 1
(1987) [Erratum {\it ibid.\/} {\bf 67}, 725 (1995)].

\bibitem{23} A. O. Caldeira and A. J. Leggett, Physica {\bf 121A},
587 (1983).

\bibitem{24} R. P. Feynman and A. R. Hibbs,
{\sl Quantum Mechanics and Path Integrals} 
(McGraw-Hill, NY, 1965).

\bibitem{25} G. W. Ford, M. Kac and P. Mazur,
J. Math. Phys. {\bf 6}, 504 (1965).

\bibitem{26} A. J. Bray and M. A. Moore, Phys. Rev. Lett.
{\bf 49}, 1546 (1982).

\bibitem{27} W. H. Louisell, {\sl Quantum Statistical
Properties of Radiation\/} (Wiley, NY, 1973).

\bibitem{28} A. Ekert and R. Jozsa, Rev. Mod. Phys. {\bf 68},
733 (1996).

\bibitem{29} D. P. DiVincenzo, Science {\bf 270}, 255 (1995).

\end{thebibliography}
\end{document}